\documentclass{aa}

\usepackage[varg]{txfonts}
\usepackage{graphicx}
\usepackage{amsmath}
\usepackage[colorlinks=true,linkcolor=magenta,citecolor=blue,filecolor=cyan,urlcolor=red,final=true]{hyperref}

\usepackage[switch,columnwise]{lineno}

\DeclareUnicodeCharacter{2212}{-}
\makeatletter
\renewcommand*\aa@pageof{, page \thepage{} of \pageref*{LastPage}}
\makeatother

\begin{document} 

   \title{Connecting remote and in situ observations of shock-accelerated electrons associated with a coronal mass ejection}
	\titlerunning{Remote and in situ observations of shock-accelerated electrons}

   \author{D.~E.~Morosan \inst{1,2}
           \and
        J. Pomoell \inst{1}
        \and
        C.~Palmroos \inst{2}
        \and
        N. Dresing \inst{2}
        \and
        E. Asvestari \inst{1}
        \and
        R. Vainio \inst{2}
        \and
        E.~K.~J.~Kilpua \inst{1} 
        \and
        J. Gieseler \inst{2} 
        \and
        A.~Kumari \inst{1,3}
        \and
        I.~C.~Jebaraj \inst{2}
        }

   \institute{Department of Physics, University of Helsinki, P.O. Box 64, FI-00014 Helsinki, Finland \\
              \email{diana.morosan@helsinki.fi}
        \and
             Department of Physics and Astronomy, University of Turku, 20014, Turku, Finland
        \and
            NASA Goddard Space Flight Center, Greenbelt, MD 20771, USA
             }

   \date{Received ; accepted }

 
  \abstract
    {One of the most prominent sources for energetic particles in our solar system are huge eruptions of magnetised plasma from the Sun called coronal mass ejections (CMEs), which usually drive shocks that accelerate charged particles up to relativistic energies. In particular, energetic electron beams can generate radio bursts through the plasma emission mechanism, for example, type II and accompanying herringbone bursts.}
    {Here, we investigate the acceleration location, escape, and propagation directions of various electron beams in the solar corona and compare them to the arrival of electrons at spacecraft.}
    {To track energetic electron beams, we use a synthesis of remote and direct observations combined with coronal modelling. Remote observations include ground-based radio observations from the  Nan{\c c}ay Radioheliograph (NRH) combined with space-based extreme-ultraviolet and white-light observations from the Solar Dynamics Observatory (SDO), the Solar Terrestrial Relations Observatory (STEREO) and Solar Orbiter (SolO). We also use direct observations of energetic electrons from the STEREO and Wind spacecraft. These observations are then combined with a three-dimensional (3D) representation of the electron acceleration locations that combined with results from magneto-hydrodynamic (MHD) models of the solar corona is used to investigate the origin and link of electrons observed remotely at the Sun to in situ electrons. }
    {We observed a type II radio burst followed by herringbone bursts that show single-frequency movement through time in NRH images. The movement of the type II burst and herringbone radio sources seems to be influenced by the regions in the corona where the CME is more capable of driving a shock. We found two clear distinct regions where electrons are accelerated in the low corona, and we found spectral differences between the radio emission generated in these regions. We also found similar inferred injection times of near-relativistic electrons at spacecraft to the emission time of the type II and herringbone bursts. However, only the herringbone bursts propagate in a direction where the shock encounters open magnetic field lines that are likely magnetically connected to the same spacecraft.}
    {Our results indicate that if the in situ electrons are indeed shock-accelerated, the most likely origin of the first arriving in situ electrons is located near the acceleration site of herringbone electrons. This is the only region during the early evolution of the shock where there is clear evidence of electron acceleration and intersection of the shock with open field lines that can connect directly to the observing spacecraft. }


\maketitle


\section{Introduction}

{Energetic particles in the heliosphere are often produced during solar flares and coronal mass ejections (CMEs) from the Sun. These particles can be detected remotely as electromagnetic emission, which they generate through various processes, or in situ by spacecraft monitoring the Sun and heliosphere. For example, energetic electrons can generate solar radio bursts through the plasma emission mechanism that can be detected remotely from ground or space-based observatories. However, it is still unknown if the electrons generating remote radio emission and all the energetic electrons observed in situ have a common origin or if there is a possible link between these two phenomena. }

{A significant amount of energetic particles from the Sun can be accelerated by collisionless plasma shocks driven by CMEs. Fast electron beams (with energies ranging from a few to tens of keVs) accelerated by these shocks can be detected remotely as bursts of radiation at radio wavelengths produced by the plasma emission mechanism \citep[e.g.,][]{wild1950, nelson1985, kl02}. The radio signatures of shock accelerated electrons are classified as type II radio bursts \citep[e.g.,][]{ma96,ne85}, while the signatures of individual electron beams escaping the shock are observed as herringbone radio bursts in dynamic spectra \citep[e.g.][]{holman1983,ca87,mo19a}. Type II bursts drift from high to low frequencies in dynamic spectra and consist of lanes with a 2:1 frequency ratio representing emission at the fundamental and harmonic of the plasma frequency. Type II bursts are associated with expanding CME-driven shock waves \citep[e.g.,][]{zimovets2012, zucca2014, zu18, mancuso19, Morosan2020a} and in rare cases with the expansion of a coronal shock wave in the absence of a CME \citep[e.g.,][]{magdalenic12, su2015, maguire2021, morosan2023}. Type II bursts often show split-bands of emission that are believed to originate from different regions of the shock front \citep[e.g.,][]{holman1983, bhunia2023} or from the upstream and downstream regions of the shock \cite[e.g.,][]{Smerd1975, kumari2017b}. Herringbone bursts usually stem from a type II lane or `backbone', however, they are sometimes observed without an accompanying type II burst \citep[][]{holman1983,ca89,mann05,mo19a}. Herringbones appear as fast drifting lanes in dynamic spectra that can drift to both high and low frequencies and represent signatures of individual electron beams escaping the shock, sometimes in opposite directions \citep{zl93, ca13, mo19a}. Unlike the backbone of type II bursts, herringbones have only been imaged on rare occasions \citep[][]{ca13, mo19a, morosan2022} and were found to closely follow the propagation of extreme-ultraviolet (EUV) waves in the low corona and fast lateral expansion of the CME at higher altitudes. The EUV wave represents a fast-mode wave or shock wave that propagates in the low solar corona ahead of the expanding CME  \citep{long2008, kienreich2009, warmuth2015}. \citet{morosan2022} suggested that the close association of herringbones with the EUV wave and CME expansion pinpoints to the formation of a large-amplitude wave that undergoes steepening into a shock during the onset of the CME eruption.  Furthermore, \citet{morosan2022} showed herringbone bursts emanating from regions of exclusively closed field lines during the early stages of the CME eruption. Such observations demonstrate that electrons generating the herringbones in the low corona do not necessarily propagate into interplanetary space but may be confined to the low corona. Therefore, for a majority of herringbone events, it is not known where the shock-accelerated electron beams propagate and under what conditions they escape the low corona. The consecutive question that naturally arises in the case of their escape is their connection to the ones observed in situ at spacecraft, which remains unclear, particularly concerning electrons with energies reaching up to a few MeV. As previously demonstrated in \citet{morosan2022}, a novel understanding of the aforementioned may be developed with the use of radio observations and in particular, the imaging of herringbones at wavelengths corresponding to the low corona. }


\begin{figure}[ht]
    \centering
    \includegraphics[width=0.99\linewidth]{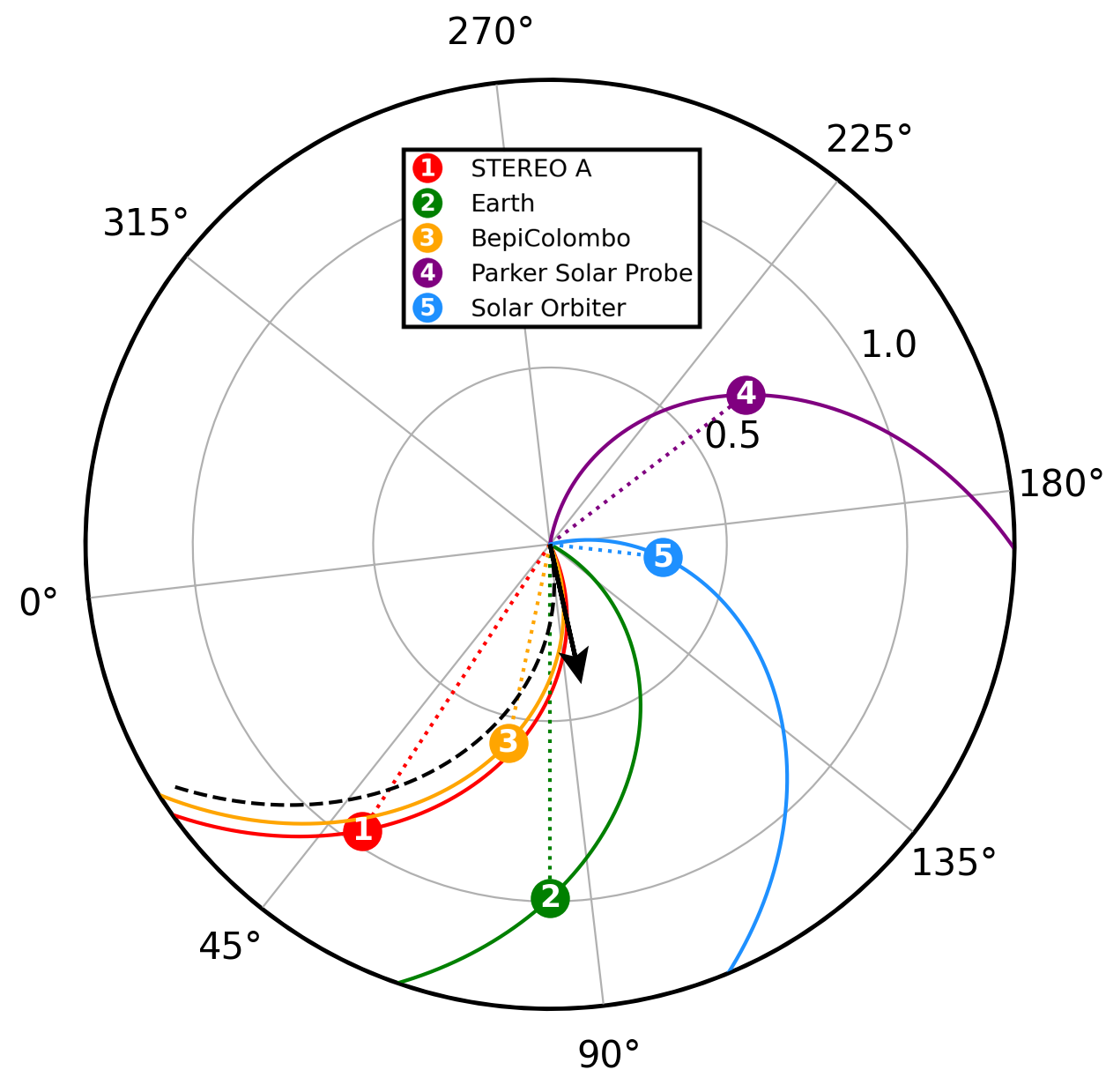}
    \caption{The location of relevant spacecraft monitoring the Sun on 28 March 2022 at 11:30~UT. The Solar-MACH plot \citep[][]{gieseler2023} shows the location and connectivity to the Sun of STEREO-A, Earth, BepiColombo, PSP and SolO using a solar wind speed of 400~km/s. The black arrow denotes the propagation direction of the CME.}
    \label{fig:fig0}
\end{figure}

{Energetic electrons are often observed directly by spacecraft with typical energies ranging from keVs up to a few MeVs. 
While solar flares were originally considered the main source of solar energetic electrons \citep{lin1982, Reames1999}, type II radio bursts provide proof of shock-accelerated electrons in the energy range of a few to tens of keV \citep[][]{mann05}. However, the acceleration of electrons up to MeV energies remains elusive and strongly debated \citep[e.g.,][]{Kahler2007, Klein_Dalla2017, Dresing2020, jebaraj2023}. A surprising result of a statistical analysis comparing peak intensities of solar energetic electron events with coronal shock parameters at the location of the magnetic footpoint of the observing spacecraft \citep{Dresing2022} was that MeV electrons showed higher correlations with shock parameters than $<100$~keV electrons. Statistical studies have also shown associations between solar energetic particles (SEPs) and solar radio bursts \citep[e.g.,][]{Kahler2019}. However, distinguishing the association of either type III or type II radio bursts with SEP events has been challenging \citep[e.g.,][]{gopalswamy2006coronal, Kahler2007, Kahler2019}. A direct association has not yet been achieved, with most studies relying on co-temporal occurrences in order to make a link. For example, \citet{klassen2011} found a series of type III radio bursts to be temporally associated with a series of electron spikes in the low energy range (below 120~keV) detected by spacecraft, these two phenomena in turn co-temporal with coronal jets. A more recent study by \citet{dresing23} used multi-spacecraft radio observations to infer the presence of a series of distinct SEP injections with significantly different propagation directions, which formed a complex widespread multi-spacecraft SEP event. Another recent study by \citet{jebaraj2023} showed both a flare and shock contribution to the acceleration of relativistic electrons, with the shock contribution supported by the presence of herringbone bursts. Up until recently, due to the limited radio imaging observations available during SEP events, especially imaging of herringbones, and the few spacecraft within 1~AU monitoring the Sun, a spatial connection to the trajectory of radio emission and the injection sites of electrons inferred from spacecraft locations could not be investigated in detail. }

{In this paper, we present the first detailed radio imaging observations of a type II burst followed by herringbones and compare their onset times and locations to those of energetic electrons observed by spacecraft. In Sect.~\ref{sec:analysis}, we give an overview of the observations and data analysis techniques used. In Sect.~\ref{sec:results}, we present the results, which are further discussed in Sect.~\ref{sec:discussion}, while our conclusions are presented in Sect.~\ref{sec:conclusion}.}


\section{Observations and data analysis} \label{sec:analysis}

\subsection{Coronal observations}

\begin{figure*}[ht]
    \centering
    \includegraphics[width=\linewidth]{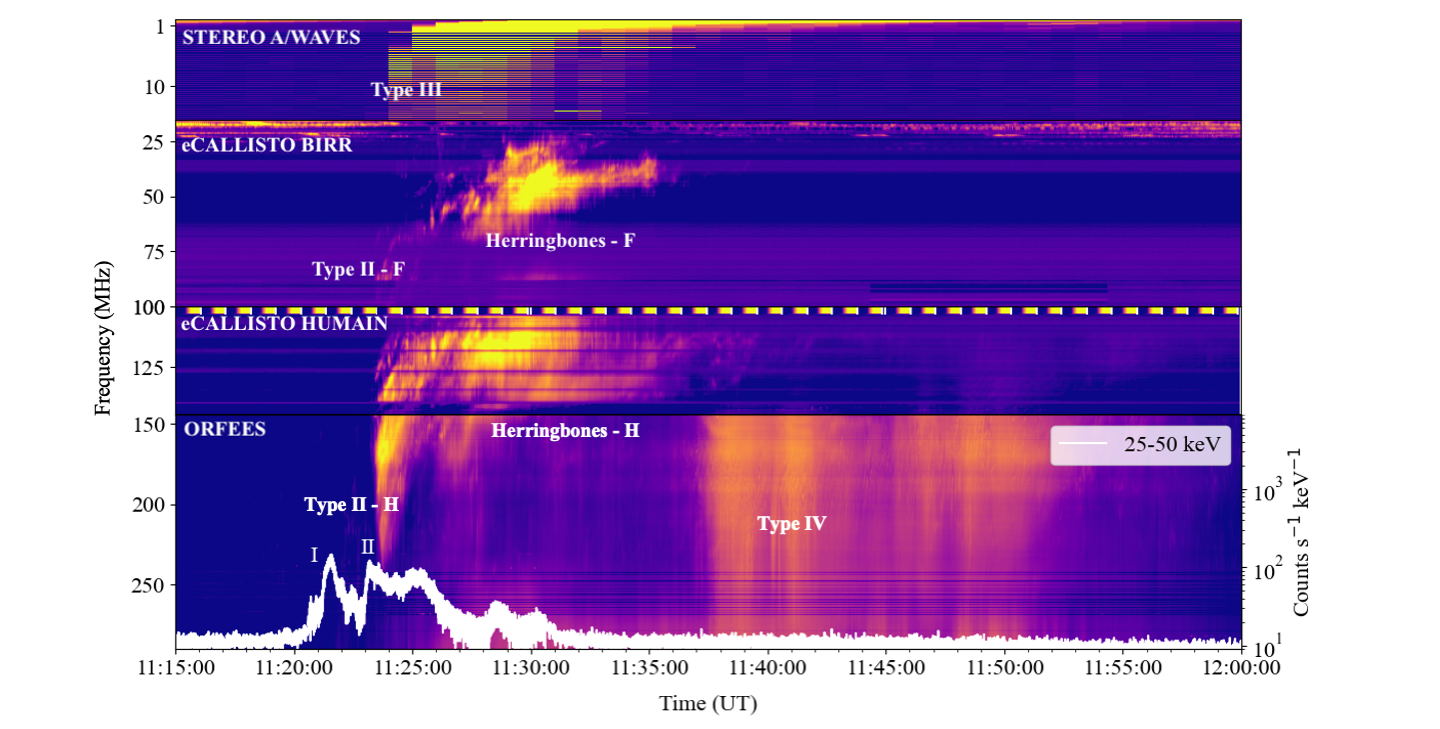}
    \caption{Composite dynamic spectrum of a complex radio event at low frequencies (1--260~MHz) on 28 March 2022. The dynamic spectrum consists of spectra from STEREO-A/WAVES, eCALLISTO Birr, eCALLISTO Humain and ORFEES. The labels 'F' and 'H' refer to fundamental and harmonic emission, respectively. The white light-curve overlaid on the ORFEES spectrum represents the Fermi GBM HXR count rate at 25--50~keV.}
    \label{fig:fig1}
\end{figure*}

{On 28 March 2022, a complex radio event was observed simultaneously with the onset of a fast CME originating from an active region close to the central meridian. This event was associated with an M4-class solar flare that began at 10:58 UT. The flare and CME were observed in remote-sensing observations by multiple spacecraft such as the Solar and Heliospheric Observatory \citep[SOHO;][]{do95, br95}, Solar Dynamics Observatory \citep[SDO;][]{pe12} and Solar Terrestrial Relations Observatory \citep[STEREO;][]{ka08} located in orbits near Earth and Solar Orbiter \citep[SolO;][]{muller2020}. SolO was located approximately in quadrature to Earth and STEREO-A but at a closer distance to the Sun (0.3~AU). The spacecraft fleet observing the Sun at the time can be seen in the Solar-MACH plot \citep{gieseler2023} in Fig.~\ref{fig:fig0} together with the CME direction (obtained from multi-viewpoint reconstructions) represented by a black arrow. Fig.~\ref{fig:fig0} shows a good nominal connectivity of STEREO-A and BepiColombo to the event (see the dashed black spiral). }

\begin{figure*}[ht]
\centering
    \includegraphics[width=0.88\linewidth]{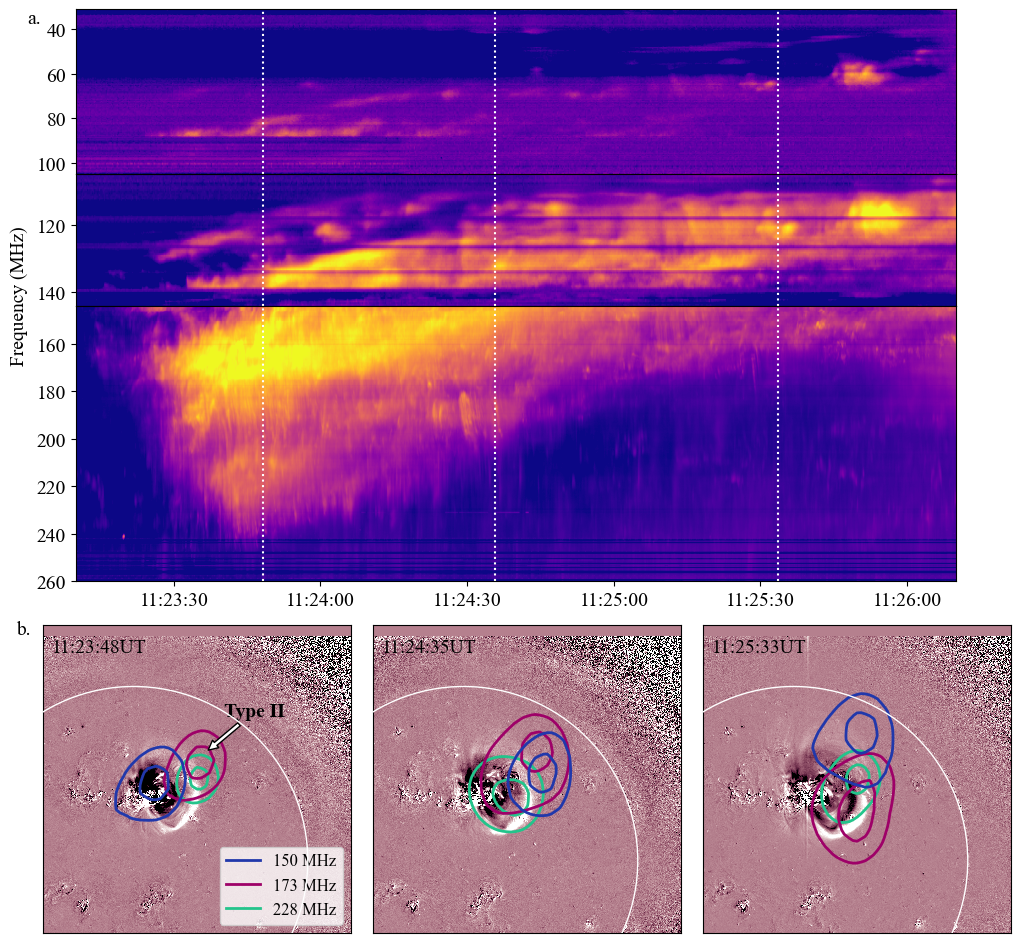}
    \caption{The type II burst in dynamic spectra and radio images. a. Combined ORFEES and eCALLISTO dynamic spectrum during the type II burst. b. Imaging of the type II radio sources overlaid as radio contours on AIA 211~{\AA} running difference images showing the evolution of the CME eruption. The radio contours are shown at three NRH frequencies that cover the extent of the type II bands: 150~MHz (blue), 173~MHz (red) and 228~MHz (green). The radio contours represent 40 and 80\% of the maximum intensity level in each image.}
    \label{fig:fig2}
\end{figure*}

\begin{figure*}[ht]
\centering
    \includegraphics[width=0.88\linewidth]{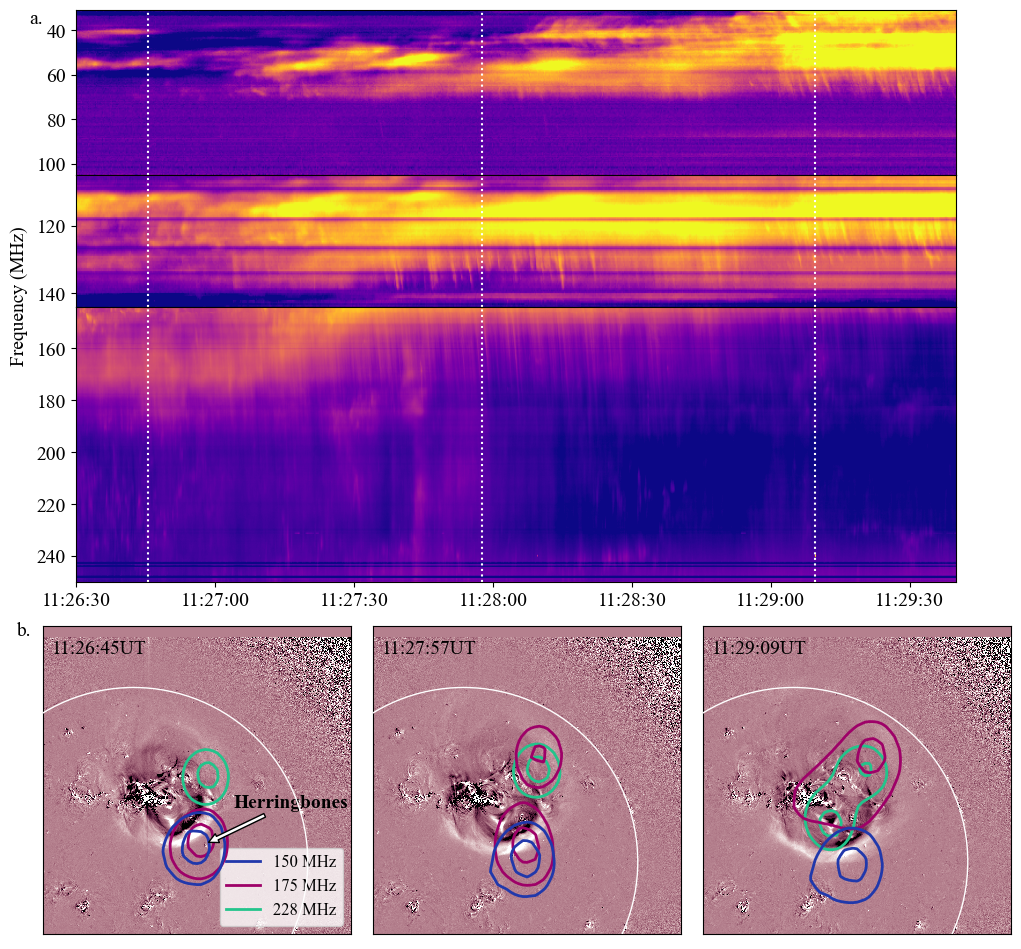}
    \caption{Herringbones in dynamic spectra and radio images. a. Combined ORFEES and eCALLISTO dynamic spectrum during the herringbone bursts following the type II in Fig.~\ref{fig:fig2}. b. Imaging of the herringbone radio sources overlaid as radio contours on AIA 211~\AA running difference images showing the evolution of the CME eruption. The radio contours are shown at three NRH frequencies: 150~MHz (blue), 173~MHz (red) and 228~MHz (green). The herringbones are only observed at 150~MHz and 173~MHz. The radio contours represent 40 and 80\% of the maximum intensity level in each image.}
    \label{fig:fig3}
\end{figure*}

{The radio event can be seen in the composite dynamic spectrum in Fig.~\ref{fig:fig1} from STEREO-A/WAVES \citep{bougeret200B}, e-CALLISTO Birr \citep{zu12}, e-CALLISTO Humain and Observation Radio pour FEDOME et l'{\'E}tude des {\'E}ruptions Solaires \citep[ORFEES;][]{Hamini2021}. The radio event starts with a type II burst at 11:23~UT that consists of both fundamental and harmonic emission lanes with band splitting labelled in Fig.~\ref{fig:fig1} and shown in more detail in the zoomed in region in Fig.~\ref{fig:fig2}. Following the initial type II burst, there are other type II-like emission lanes that consist of clear herringbone bursts which are labelled as 'Herringbones' in Fig.~\ref{fig:fig1} and shown in more detail in the zoomed in region in Fig.~\ref{fig:fig3}. Following the type II and herringbones, a type IV continuum is visible in the ORFEES dynamic spectrum. At lower frequencies (< 16~MHz) a prominent group of type III bursts was observed from space by the STEREO-A/WAVES and Wind/Waves instruments, with an onset time slightly after the type II burst at 11:24~UT. The type III burst is not visible at higher frequencies in the available spectra.}

{Hard X-rays (HXRs) were also associated with this event and observations are available from the Fermi Gamma-ray Burst Monitor \citep[GBM;][]{fermi2009}. The 25--50~keV light-curve is overlaid on the ORFEES dynamic spectrum in Fig.~\ref{fig:fig1}. The light-curve shows two episodes of HXR acceleration that are labelled as I and II in Fig.~\ref{fig:fig1}. The onset of HXRs in this energy range is 11:20~UT at the spacecraft and a similar onset time is observed in the 50--100~keV energy range. Beyond these energies, no HXRs are observed. A comprehensive analysis of the X-ray observations associated with this event with the  Spectrometer Telescope for Imaging X-rays (STIX) onboard SolO is presented by \citet{purkhart2023}. For the present study, we are mainly interested in the onset time of HXRs at the Sun. }

{The harmonic emission of the type II burst (labelled as `Type II - H' in Fig.~\ref{fig:fig1}) and herringbones (labelled as `Herrinbones - H\ in Fig.~\ref{fig:fig1}) can be imaged by the Nan{\c c}ay Radioheliograph \citep[NRH;][]{ke97} in the following frequency bands: 150, 173 and 228~MHz. Contours of the radio emission sources are shown at three frequencies (150--blue, 173--red, 228--green) in Figs.~\ref{fig:fig2} and \ref{fig:fig3}, overlaid on running difference images from the Atmospheric Imaging Assembly (AIA) onboard SDO \citep{le12} at a wavelength of 211~\AA, that show the evolution of the CME eruption and associated shock wave. The locations of the radio bursts at different frequencies can also be seen in Movies 1, 2 and 3 accompanying this paper, which present a full evolution of the eruption at EUV and radio wavelengths. The centroids of the radio bursts were extracted using the weighted mean method to determine the movement of the radio sources through time (Figs.~\ref{fig:fig5} and \ref{fig:fig6}). }

{The onset and propagation of the eruption and its associated shock wave can be seen in running difference images from AIA (Figs.~\ref{fig:fig2} and \ref{fig:fig3}) and Movie 1 accompanying this paper. The eruption in the low corona is clearly visible in extreme ultraviolet (EUV) images from three vantage points: Earth (SDO, SOHO), STEREO-A, and SolO (Fig.~\ref{fig:fig7}). The first observation of the CME in white-light images was at 11:30~UT in the STEREO-A’s inner coronagraph COR1 \citep{ho08}. A prominent EUV wave was also observed during the eruption at EUV wavelengths from SDO/AIA and STEREO-A's Extreme Ultraviolet Imager \citep[EUVI;][]{ho08}.  }

\subsection{Spacecraft particle observations}

{Energetic electrons and other particles from the Sun were monitored by the spacecraft constellation shown in Fig.~\ref{fig:fig0}. A magnetic connectivity analysis indicates a strong magnetic connection between the eruptive region and the Parker spiral magnetic field lines connecting to BepiColombo and STEREO-A. Energetic particle measurements with reliable statistics for this event are only available from STEREO-A, Wind, and SOHO. BepiColombo particle observations are unfortunately not available as the Solar Intensity X-ray and Particle Spectrometer \citep[SIXS; ][]{sixs2020} and the Environment Radiation Monitor \citep[BERM;][]{berm2022} did not have sufficient statistics to accurately determine the onset times of energetic particles. Consequently, we were unable to utilize in-situ particle measurements from BepiColombo for this event.}

{Energetic electrons are first observed at STEREO-A by the Solar Electron and Proton Telescope \citep[SEPT;][]{Muller-Mellin2008} instrument covering near-relativistic energies and by the High Energy Telescope \citep[HET; ][]{vonRosenvinge2008} measuring relativistic electrons. Near Earth, Wind/3DP \citep[Wind, 3DP:][]{Ogilvie1997, Lin1995} and SOHO's Electron Proton Helium Instrument \citep[EPHIN;][]{MuellerMellin1995} observe the first electrons $\sim$10 minutes later than STEREO-A. The electron event as observed by STEREO-A/SEPT and Wind/3DP is shown in Fig.~\ref{fig:fig8}. The first high-energy protons observed by STEREO-A/HET and SOHO's Energetic and Relativistic Nuclei and Electron experiment \citep[ERNE;][]{Valtonen1997} arrive $\sim$10 minutes later than the first electrons at respective observers. }

\section{Results} \label{sec:results}

\begin{figure*}[ht]
\centering
    \includegraphics[width=0.9\linewidth]{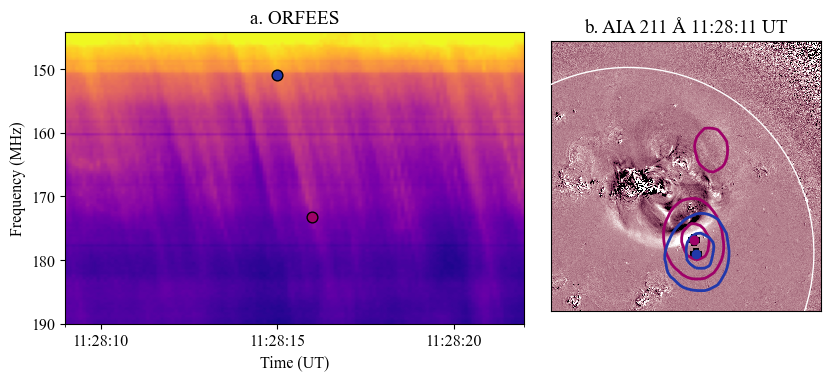}
    \caption{Zoom-in of the reverse drift herringbone together with their location at two frequencies. a. The reverse drift herringbones are shown in a zoomed in dynamic spectrum from ORFEES and can be imaged by the NRH at 150~MHz (blue dot) and 173~MHz (magenta dot). b. Radio contours at 40 and 80\% of the maximum intensity level at the time and frequency denoted by the blue and magenta dots in (a) overlaid on an AIA 211~A running difference image. The centroids of these radio contours are also shown as plus symbols with the corresponding colour.}
    \label{fig:fig4}
\end{figure*}

\begin{figure*}[ht]
\centering
    \includegraphics[width=\linewidth]{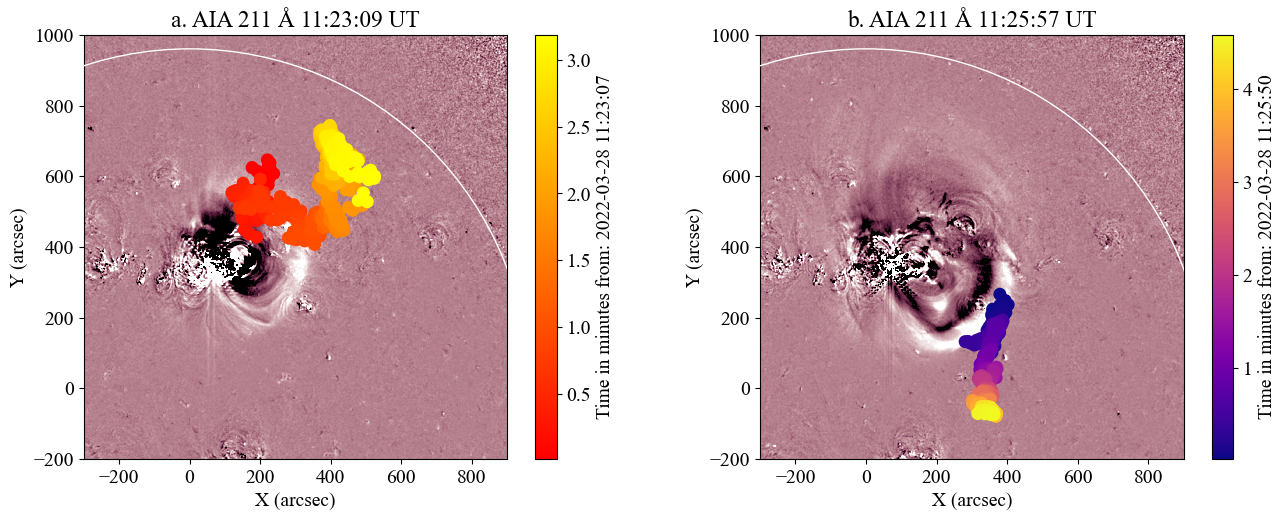}
    \caption{Centroids of the type II (left) and herringbone (right) locations obtained from NRH images at 150~MHz. The centroids are colour coded and the colour represent time in minutes from the onset time of each emission. }
    \label{fig:fig5}
\end{figure*}

\subsection{Location and characteristics of the radio emission }

{The type II and herringbone bursts presented in the dynamic spectrum in Fig.~\ref{fig:fig1} are observed over a wide range of frequencies (25--250~MHz) for a duration of $\sim$15~minutes. Imaging observations from the NRH (Figs.~\ref{fig:fig2} and \ref{fig:fig3}) show that the type II and subsequent lanes composed of herringbones are not part of the same radio burst as emission originates from different locations. The association between the bursts observed in ORFEES to the radio sources in the NRH images was done by comparing the intensity of the radio bursts in the ORFEES data to the integrated flux over the extent of individual radio sources in NRH images (for more details see Appendix A). The radio sources that are not labelled in Figs.~\ref{fig:fig2} and \ref{fig:fig3} are likely to correspond to the continuum emission associated with this event. The type II occurs north of the eruption and follows the CME flank and EUV wave in a north-westerly direction (Figs.~\ref{fig:fig2}), while the herringbones originate south of the eruption site and propagate with the EUV wave southwards. Thus, we will treat these bursts separately throughout the paper and refer to them as type II and herringbone lanes, respectively. We note that the type II also consists of fine structures including herringbones. However, the type II fine structures stem from a more sloped backbone, while the later herringbone bursts appear to stem from a flat backbone. The herringbone lanes also have better defined and longer bandwidth herringbones such as the ones shown in Figs.~\ref{fig:fig3} and \ref{fig:fig4}.  }

{The imaging observations show that there are at least two different locations at the shock front where electrons are accelerated. Both of these regions correspond to the expanding EUV wave front in the low corona visible in the AIA running difference images in Figs.~\ref{fig:fig2} and \ref{fig:fig3}. The type II and herringbones propagate slightly ahead of the EUV wave in plane-of-sky images and evolve co-temporally following the same outward expansion of the wave. This has been observed before in previous studies where a close association between type II, herringbones and coronal waves have been recorded \citep[e.g.,][]{uchida1974, vrsnak2005, ca13, mo19a, morosan2022, morosan2023}. }

{The NRH images offer a good coverage of the reverse drifting herringbones over time at two frequencies: 150 and 173~MHz. A zoom-in of these reverse drift herringbones is shown in Fig.~\ref{fig:fig4}. One such herringbone is also imaged in Fig.~\ref{fig:fig4} at these two frequencies. There is a spatial separation between the centroids that is $\sim$54~Mm in the plane-of-sky. Using this separation and the duration of the herringbone emission we can estimate a speed of 0.18~\textit{c}, which represents the speed of the electron beam producing this emission. This is comparable to the deduced speed from previous herringbone observations of 0.16~\textit{c} \citep[e.g.,][]{mann05, ca15}. The energy of the emitting electrons estimated from their speed is $\sim$8.4~keV. We note, however, that these values are obtained using the plane-of-sky positions of the herringbone centroids and thus they are a lower limit due to likely projection effects.}

{The centroids of the type II and herringbone sources over time are shown in Fig.~\ref{fig:fig5} at 150~MHz. The colouring of the centroids represents time in minutes in each of the panels. The type II centroids show an unusual curved trajectory travelling in a north-westerly direction from the eruption site. The herringbones have a straight trajectory propagating southwards from the eruption region. These propagation directions are similar at 173~MHz for both the type II and herringbones, respectively. Thus, there are two distinct regions where electrons are accelerated during the CME eruption.}

\begin{figure*}[ht]
\centering
    \includegraphics[width=\linewidth]{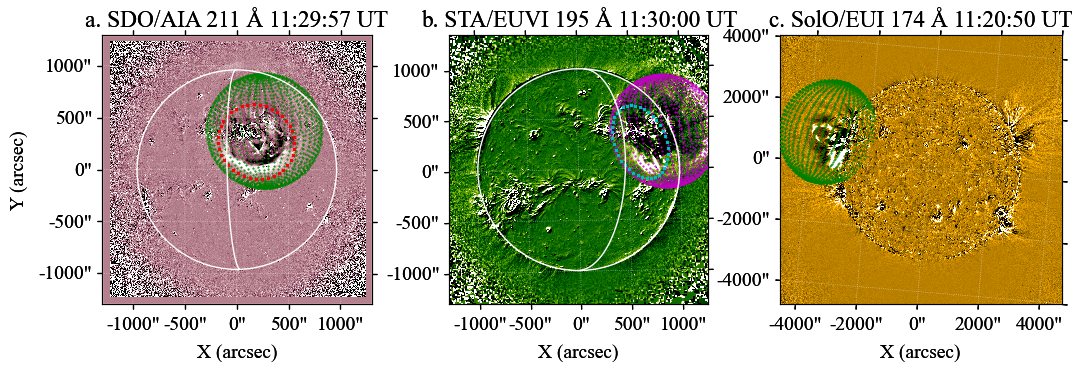}
    \caption{Shock reconstruction in 3D using multiple view-points. The reconstruction at $\sim$11:30~UT is shown as a wiremesh overlaid on EUV running-differences images of the Sun from three perspectives: SDO, STEREO-A and SolO. The locations of these spacecraft is shown in Fig.~\ref{fig:fig0}. The CME eruption without the overlapping wiremesh can also be seen in the supplementary movies accompanying the paper. The time stamps in the title of each panel represent times at each individual spacecraft. The time at SolO is four minutes earlier compared to STEREO-A and Earth after correcting for light-travel time. The visible solar limb as viewed from SolO is overlaid in the SDO and STEREO-A images as a white arc.}  
    \label{fig:fig6}
\end{figure*}

\begin{table*}[ht]
\centering
\caption{Spheroid fitting parameters determined using the PyThea tool.}
\label{tab:spheroid_param}
\begin{tabular}{|c||c|c|c|c|c|c|c|c|}
\hline
Time (UT) & Heliographic & Heliographic& heliocentric & radial & orthogonal & apex & self-similar & eccentricity \\
 & longitude & latitude & distance $r_{c}$ & axis & axis & height & constant $\kappa$ & $\epsilon$\\
\hline
\hline
11:25:00 & 12.8 & 17.7 & 1.16 & 0.31 & 0.33 & 1.47 & 0.71 & -0.39\\
11:27:30 & 12.8 & 17.7 & 1.21 & 0.41 & 0.44 & 1.62 & 0.71 & -0.39\\
11:30:00 & 12.8 & 17.7 & 1.28 & 0.54 & 0.58 & 1.82 & 0.71 & -0.39\\
11:32:30 & 12.8 & 17.7 & 1.35 & 0.66 & 0.72 & 2.01 & 0.71 & -0.39\\
11:35:00 & 12.8 & 17.7 & 1.41 & 0.78 & 0.84 & 2.19 & 0.71 & -0.39\\
11:37:30 & 12.8 & 17.7 & 1.48 & 0.90 & 0.97 & 2.38 & 0.71 & -0.39\\
\hline

\end{tabular}
\end{table*} 

\subsection{The coronal mass ejection and radio emission in three dimensions}
 
{Signatures associated with the eruption were observed in EUV images (EUV wave \& ejected plasma) and white-light imaging (CME \& leading shock) from the vantage points of Earth, STEREO-A and SolO (Fig.~\ref{fig:fig6}). These three perspectives allow us to reconstruct the shock wave in 3D by fitting a spheroid to all vantage points simultaneously, aiming for the most optimal fit. This analysis was performed using the PyThea CME and shock wave analysis tool \citep{Kouloumvakos2022}. The shock reconstruction was applied to six time steps on 28 March 2022, while the shock is visible in EUV filtergrams. Since the spheroid is an idealised shape and shock waves are far more complex in reality, fitting all visible characteristics in EUV filtergrams was not possible, as can be seen in Fig.~\ref{fig:fig6}. Namely, the spheroid parameters selected to fit the well defined dome shape of the shock above the solar surface, as observed in STEREO-A/EUVI at 195~$\AA$, resulted in a spheroid mesh extending beyond the shock signatures observed in SDO/AIA at 211~$\AA$, except for the northern flank where the shock is visible outside the mesh (Fig.~\ref{fig:fig6}a and Supplementary Movie 2). Furthermore, in SolO/EUI at 174~$\AA$, the shock shape observed strays away from the spheroid shape, showing signatures of bulging at its northern front which therefore cannot be fitted well by the spheroid mesh. To better see these signatures please see Supplementary Movie 3 accompanying the paper which follows the CME evolution in SolO. The fitted parameters are given in Table~\ref{tab:spheroid_param}. These are the heliographic longitude and latitude, and the heliocentric distance  of the spheroid center $r_{c}$ (columns 2, 3~\&~4), the radial and orthogonal axis of the spheroid (columns 5~\&~6), the height of the apex of the spheroid (column 7), the self-similar constant $\kappa$ (column 8) defined as the ratio of one of the semi-axis and the height of the apex \citep[see][for definitions]{Kouloumvakos2022}, and the eccentricity $\epsilon$ of the spheroid (column 9). As can be seen from the table the radial and orthogonal axis are close in value, as supported by the dome shape seen in SDO/AIA at 211~$\AA$ and STEREO-A/EUVI at 175~$\AA$ early on during the eruption. Based on the fitted values of the apex height as well as the radial and orthogonal axes, the shock appears to be expanding with an initial acceleration between 11:25~UT to 11:30~UT before decelerating within the next 2~min of evolution and settling to a constant speed of $\sim$560~km/s for the expansion along the radial axis of the spheroid, $\sim$610~km/s along the orthogonal axis of the spheroid, and $\sim$870~km/s for the expansion of the apex. The maximum expansion speeds reached are $\sim$600~km/s, $\sim$830~km/s, and $\sim$930~km/s along the radial and orthogonal axis of the spheroid, and of the apex respectively. The herringbone bursts occur during this peak acceleration period (for more information see Fig.~B.1 in Appendix~B that shows the shock expansion through time).}

\begin{figure*}[ht!]
\sidecaption
\centering
    \includegraphics[width=0.7\linewidth, trim = 1.6cm 0cm 1.6cm 0cm, angle = 0]{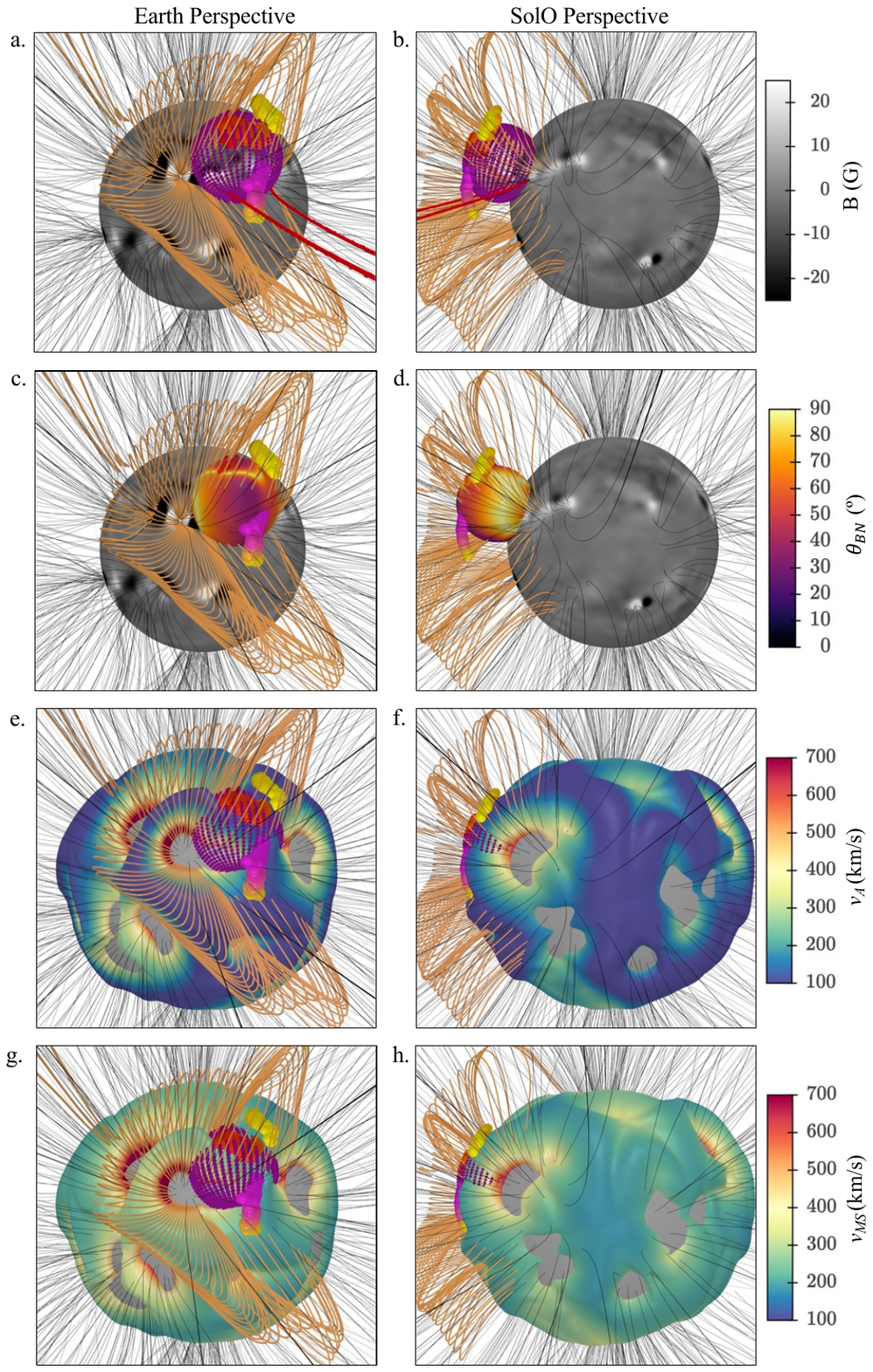}
    \caption{Coronal model results combined with the shock reconstruction and radio emission locations from two perspectives: Earth (left) and SolO (right). \textit{a--b.} Photospheric magnetogram overlaid with MAST open (black) and closed magnetic field lines (orange), the shock reconstruction at 11:25~UT observation time at Earth (magenta wiremesh) and centroids of type II and herringbone radio sources (coloured spheres). The red open field lines are example of field lines connecting to STEREO-A. \textit{c--d.} Photospheric magnetogram overlaid with the same features as in panels \textit{a--b.} except that the shock reconstruction is overlaid with values of $\theta_{BN}$. \textit{e--f.} Density isosurface overlaid with values of Alfv\'en speed ($v_A$) obtained from the MAST model. Overlaid on the isosurface are open (black) and closed magnetic field lines (orange), the shock reconstruction at 11:25~UT (magenta wiremesh) and centroids of type II and herringbone radio sources (coloured spheres). The grey areas represent a sphere surface at 1.2~R$_{\odot}$ to give a visual indication of the height of the density isosurface. \textit{g--h.} Density isosurface overlaid with values of the fast magnetosonic speed ($v_{FMS}$) obtained from the MAST model. Overlaid on the isosurface are the same features as in panels \textit{e--f}. For different view points please see the HTML file accompanying this paper.} 
    \label{fig:fig7}
\end{figure*}

{The shock reconstruction at $\sim$11:30~UT can be seen in Fig.~\ref{fig:fig6} as a wire mesh overlaid on each of the three view points. The 3D reconstruction of the shock at the same time can be seen in Figs.~\ref{fig:fig7}a and \ref{fig:fig7}b as a magenta wireframe from two perspectives: Earth and SolO.  }

\begin{figure}[ht]
\centering
    \includegraphics[width=\linewidth, trim = 0cm 1cm 0cm 0cm]{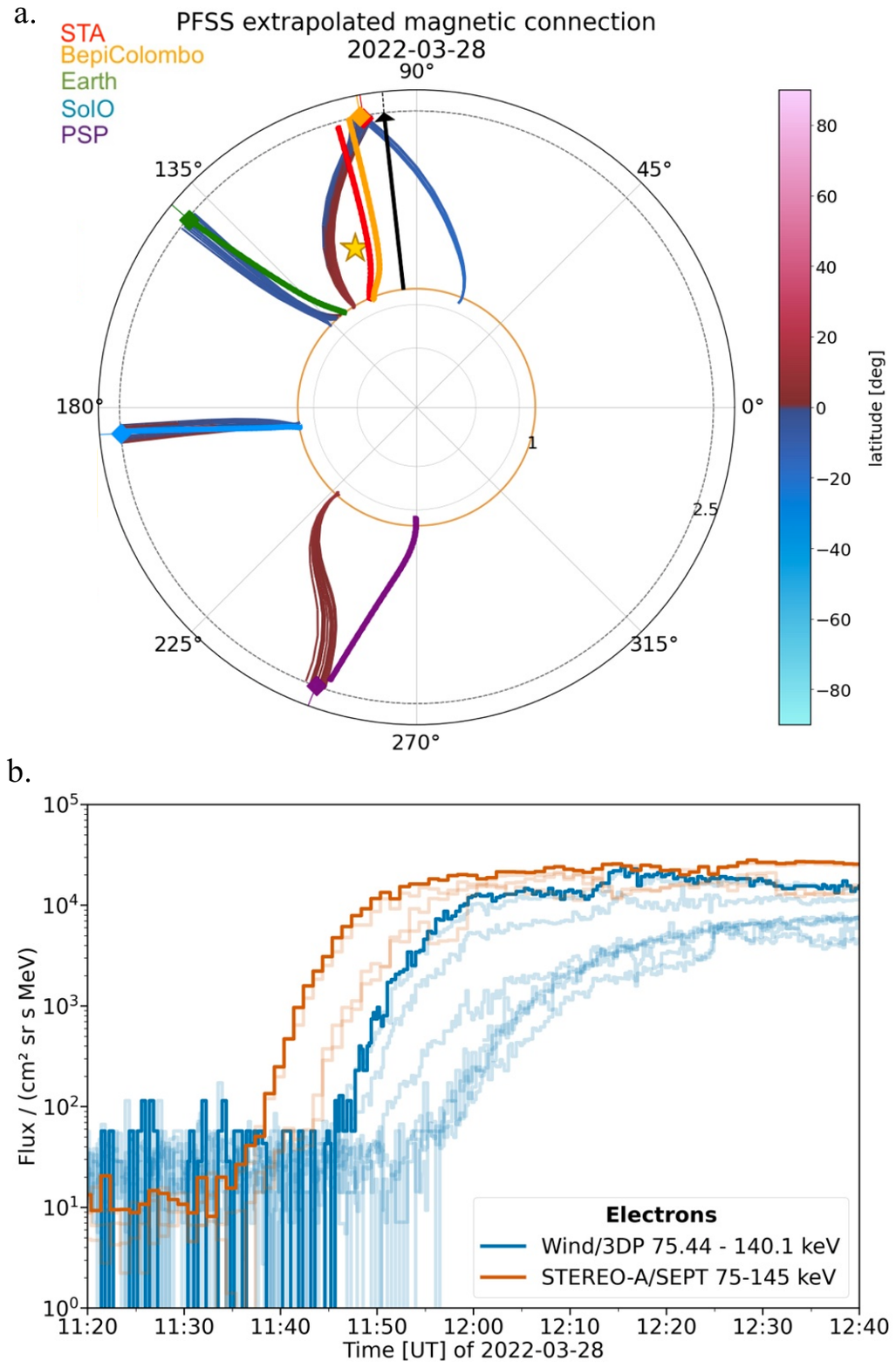}
    \caption{Magnetic connectivity of various spacecraft to the Sun and the first arriving particles at these spacecraft. a. Ecliptic plot showing the spacecraft connectivity to the source surface which is represented by coloured diamonds for STEREO-A (red), BepiColombo (yellow), Earth (green), SolO (blue) and PSP (purple) using PFSS (blue--pink) and MHD field lines (colour-coded by spacecraft). The black arrow corresponds to the direction of CME propagation. The star symbol denotes the location of one of the herringbone sources at $0^\circ$ latitude. b. Energetic electron intensities at $\sim$100~keV as observed in different viewing directions by STEREO-A/SEPT (orange curves) and Wind/3DP (blue curves) near Earth. The viewing directions observing the first arriving electrons are indicated by the stronger coloured curves.}
    \label{fig:fig8}
\end{figure}

\subsection{The propagation medium of radio generating electrons}

{The shock reconstruction can show its expansion direction in 3D and combined with the radio emission locations it can localize the particle acceleration regions in the corona. To determine the coronal properties of the shock expansion regions at the locations of the type II and herringbones, we employed the Magnetohydrodynamics Around a Sphere Thermodynamic (MAST) model \citep{lionello2009}. The MAST model is an MHD model developed by Predictive Sciences Inc.\footnote{http://www.predsci.com} and can be used to study the global structures and dynamics of the solar corona. The model can produce estimates of the electron density and magnetic field strength that are used in this study to compute the global Alfv\'en speed. This model uses as inner boundary conditions the magnetic field from photospheric magnetograms from the Heliospheric and Magnetic Imager \citep[HMI;][]{sche12} onboard SDO. The low coronal densities obtained from MAST have been scaled up by a factor of 2 to obtain more accurate values of their heights \citep{wang17}. The MAST electron densities are also used to de-project the radio centroids from the plane-of-sky view so that they can be related to the 3D expansion of the shock and the magnetic environment. The magnetic field obtained from the MAST model is also used to trace coronal magnetic field lines to larger heights.}

{The MAST model outputs used in this study are global densities, magnetic field and temperature of the coronal plasma. These properties are then used to compute the Alfv\'en ($v_A$) and fast magnetosonic speeds ($v_{FMS}$). The results of the model are presented in Fig.~\ref{fig:fig7} from two perspectives: Earth (left panels) and SolO (right panels), based on the methods of \citet{morosan2022}. Panels \textit{a--b} of Fig.~\ref{fig:fig7} show the photospheric magnetogram used as input to the MHD model overlaid with closed (orange) and open (black) magnetic field lines obtained from the MAST model. Also overlaid in these panels is the reconstructed shock wave at 11:25~UT shown as a magenta wireframe mesh. The de-projected radio centroids are plotted as coloured spheres for both the type II and herringbone sources, with the same colouring as in Fig.~\ref{fig:fig5}, which represents time in minutes. Both type II and herringbone sources are located outside the reconstructed shock wave. The type II overlaps with the shock during the early stages, however, the start time of the type II source is two minutes before the shock reconstruction in Fig.~\ref{fig:fig7}, which is at 11:25~UT. }

{The shock geometry, i.e. the angle between the shock normal ($\mathbf{\hat{n}}$) and the upstream magnetic field ($\mathbf{B}$), denoted as $\theta_{BN}$, is depicted in panels \textit{c--d} of Fig.~\ref{fig:fig7}. While these panels share similar features with panels \textit{a--b} of the same figure, they distinguish themselves by having the shock surface colour-coded. These colours represent $\theta_{BN}$ values ranging from 0 to 90$^{\circ}$. The type II centroids coincide with a narrow region on the shock surface where $\theta_{BN}\sim90^{\circ}$ (denoted by the yellow colour). A quasi-perpendicular shock is expected for the generation of type II and herringbone emission due to the shock-drift or fast-Fermi acceleration mechanism \citep[e.g.,][]{holman1983, cairns2003, mann2018, mann2022} and similar values of $\theta_{BN}$ were demonstrated in previous studies in the low corona also using MHD modelling \citep[e.g.,][]{kouloumvakos2021}. On the other hand, the herringbones occur near a region where the shock is oblique $\theta_{BN}\sim30^{\circ}$. The shock-magnetic field geometry does not change significantly during the duration of type II and herringbones at this location. This was checked using later time reconstructions over a 10 minute period where the shock remains oblique or quasi-parallel. }

{The last four panels in Fig.~\ref{fig:fig7}, \textit{e--f} and \textit{g--h}, show an electron density iso-surface corresponding to a density layer of $7\times10^7$~cm$^{-3}$ which is also the electron density that produces harmonic plasma emission at 150~MHz. This is in turn the frequency of the 3D centroids of the overlaid radio sources. This density iso-surface was used to estimate the z-component of the plane-of-sky radio centroids so that they can be de-projected from the plane-of-sky view. In panels \textit{e--f}, overlaid on the density iso-surface are the Alfv\'en speed values at the height and location of the density layer that go from blue (low $v_A\sim$ 100km/s) to red (high $v_A\sim$ 700km/s). In panels \textit{g--h}, overlaid on the density iso-surface are the fast magnetosonic speed values at the height and location of the density layer. }

\begin{figure*}[ht]
\centering
    \includegraphics[width=\linewidth, trim = 0cm 0cm 0cm 0cm]{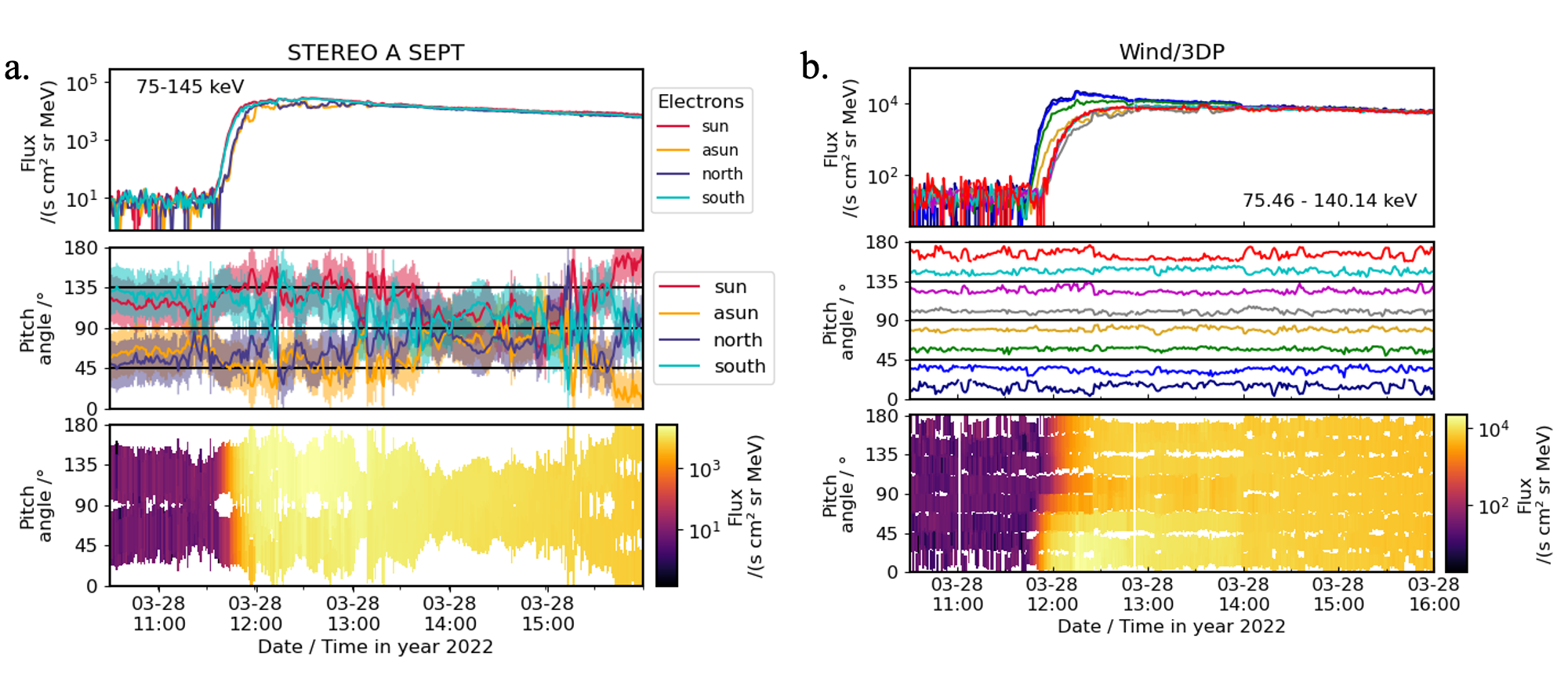}
    \caption{Near-relativistic electron measurement by STEREO-A/SEPT (a) and Wind/3DP (b). From top to bottom: Electron flux measured in different viewing directions, pitch-angles of the center of those viewing directions, and electron-pitch-angle distribution with colour coding denoting the electron flux.}
    \label{fig:fig10}
\end{figure*}

{Both the type II and herringbones radio sources are located in a region with enhanced density which forms a ridge where the Alfv\'en speed is low ($\sim$100~km/s). The fast magnetosonic speed in this region is also low  ($\sim$250~km/s). The curved trajectory of the type II can be explained by the curved shape of the density ridge it follows, as this is the only region towards the North pole where the shock can encounter low Alfv\'en and low fast magnetosonic speeds and is therefore likely to form a shock. The type II sources are located inside a region of exclusively closed magnetic field lines forming a coronal streamer in agreement with previsouly reported observations from recent studies \citep[e.g., ][]{mancuso2004, magdalenic14, frassati2019, fr20, morosan2022}. The selection of closed magnetic field lines (orange) in Fig.~\ref{fig:fig7} show part of this streamer belt. The herringbone sources propagate southward along a narrow density ridge also characterized by low Alfv\'en speeds. This ridge is located between two active regions (one of which is the active region where the eruption originates), however, it is not part of the same streamer belt as the type II burst. This region also consists of closed field lines as no open field originates from this region. This was confirmed by consulting the Potential Field Source Surface \citep[PFSS; e.g.][]{sc03} model in addition to the MAST model. Unlike the region where the type II source propagates, the herringbones are likely to encounter overlying open field lines connecting in the vicinity of the active refion of the eruption. A few of these open field lines are shown in Fig.~\ref{fig:fig7}c intersecting the herringbones as they propagate southward. }

\subsection{Electron onset and injection times and magnetic connections to spacecraft}

{The flaring active region and shock wave have a good magnetic connection to some of the spacecraft monitoring the Sun. The connection between coronal magnetic field lines and the interplanetary magnetic field connecting to various spacecraft is shown in Fig.~\ref{fig:fig8}a in Carrington coordinates. In Fig.~\ref{fig:fig8}a, the coloured solid diamonds mark the various spacecraft monitoring the Sun at the point where they magnetically connect to the source surface. The magnetic field lines connected to them from the source surface represent the nominal Parker spiral solutions computed considering their heliocentric distances and the observed solar wind speeds. Inside the dashed black circle, the magnetic connection is extrapolated with a PFSS solution \citep[][]{Stansby2020}, where the colour of the magnetic field lines corresponds to heliospheric latitude. Overlaid on the PFSS field lines are MHD field lines obtained from MAST, colour-coded based on each spacecraft. The black arrow corresponds to the direction of CME propagation. Also overlaid on this image is the location of one of the herringbone sources at $0^\circ$ latitude denoted by the star symbol. Both PFSS and MHD models show open field line connectivity to STEREO-A in the vicinity of the herringbone radio source. }

{Energetic particles were first recorded by STEREO-A followed by the Wind spacecraft at Earth. Fig.~\ref{fig:fig8}b shows that near-relativistic electrons (75--140~keV) are observed $\sim$9~minutes earlier at STEREO-A/SEPT than at Wind in a similar energy range (75--145~keV). Fig.~\ref{fig:fig10} shows the electron intensities of these energies in all viewing directions and the respective pitch-angle coverage of both STEREO-A/SEPT and Wind/3DP from 10:30 to 16:00~UT. There is noticeable anisotropy in the electron intensities of both observers, especially for Wind. Since the pitch-angle coverage of SEPT at the start of the event is limited in comparison to 3DP's, we can not rule out the possibility that the electron population is even more beamed at STEREO-A than the intensity data alone suggests. Wind observes an anisotropic rise phase with particles arriving from the solar direction (pitch angle 0) followed by a slightly bi-directional distribution with a depletion of intensities at 90 degrees pitch angle. This suggests the presence of a reflective boundary behind Wind's location leading to part of the passed electron distribution being scattered back to the spacecraft. We note that those bi-directional distributions can also be caused by a spacecraft being situated inside a magnetic cloud \citep[e.g., ][]{Dresing2016}. However, inspection of in-situ magnetic field measurements at Wind does not suggest the presence of a magnetic cloud during this period.}

\begin{table*}[t]
\centering
\caption{Onset times, inferred injection times and corresponding energies of SEPs observed by spacecraft, HXRs with various energies and electrons producing the radio bursts observed. The method for determining the injection time is shown in the last column with the inferred propagation path length $L$ in case VDA was used. The emission time at the Sun for particles is the injection time estimated using various methods presented in the last column. The emission time at the Sun in the case of the radio bursts and HXRs refers to the time when these radio bursts and HXRs are emitted taking into account the light-travel time to Earth of $\sim$8~minutes and 20~seconds. }
\label{tab:table1}
\begin{tabular}{|c||c|c|c|c|}
\hline
Event & Onset time  & Emission time  & Energy & Method for     \\
       & at spacecraft (UT) & at the Sun (UT) &   & injection time  \\
\hline
\hline
Type II             & 11:23          & 11:15               & - & -         \\
Type III            & 11:24          & 11:16                 & - & -                 \\
Herringbones        & 11:25          & 11:17               & 8~keV & -      \\
Herringbones (solar equator)        & 11:28         &  11:20              & 8~keV & -              \\
\hline
HXR I       & 11:20             & 11:12              & 25--50~keV              &-   \\
HXR II        & 11:23             & 11:15              & 25--50~keV              &-   \\
\hline
Electrons (STA/SEPT)  & 11:38            & 11:15$^{+4}_{-5}$ min   & 45-55~keV & TSA        \\
Electrons (STA/SEPT)  & 11:34            & 11:21$^{+2}_{-3}$ min   & 225--255~keV & TSA        \\
Electrons (STA/HET)   & 11:35            & 11:25$^{+2}_{-2}$ min   & 0.7--1.4~MeV & TSA      \\
Protons (STA/HET)     & 11:51(*)           & 11:26$^{+9}_{-9}$ min     & 13.6--100~(60--100*)~MeV  & VDA $L=1.170^{+0.237}_{-0.170}$~AU        \\
\hline
Electrons (Wind/3DP)   & 11:47           & 11:22$^{+7}_{-8}$ min    & 46.16--85.73~keV         & TSA  \\
Electrons (SOHO/EPHIN) & 11:46           & 11:33$^{+4}_{-5}$ min    & 0.25--0.7~MeV      & TSA       \\
Protons (SOHO/ERNE)    & 12:04 (*)       & 11:40$^{+15}_{-15}$ min    & 13--130~(64--80*)~MeV        & VDA $L=1.377^{+0.463}_{-0.377}$~AU         \\
\hline

\end{tabular}
\end{table*}

{In order to connect the in-situ SEP observations of STEREO-A and Wind with the different observed radio features, we infer the SEP injection times at the Sun. SEP onset times, inferred SEP injection times, as well as the onset time of various radio bursts are summarised in Table~\ref{tab:table1}. The injection time in the case of the solar radio bursts represent the times at which the bursts are emitted close to the Sun, taking into account the travel time of light from Sun to Earth, i.e. 8~minutes and 20~seconds. To infer the solar injection times of SEPs, we employed a velocity dispersion analysis \citep[VDA, ][]{Lintunen2004}. By assuming a common injection time and propagation path length of all particles, VDA allows inferring these two parameters through the systematic different arrival times of SEPs caused by their different energies, which is called velocity dispersion. Because an electron VDA did not yield physical results for the present event, we estimated the onset time of energetic electrons using different energy ranges of both STEREO-A and near-Earth observations and performed a time-shift analysis (TSA). TSA requires that a certain propagation path length is assumed. Then the injection time is simply calculated by backtracking this path length starting at the SEP onset time using the speed corresponding to the mean energy of the particles in a given energy channel. While the path length is often assumed to be an ideal Parker spiral that corresponds to the observer's heliocentric distance and typical solar wind speed of 400 km/s, in this study we acquired the path length by performing a VDA for protons at the respective vantage points using STEREO-A/HET and SOHO/ERNE observations. The VDA provides the path length and injection time accompanied by error bars which are symmetrical. We correct these error bars by taking into account that the path length can not, however, be shorter than the radial distance between the Sun and the spacecraft. Hence the uncertainty of the path length is bounded from below by the radial distance, which is $\sim$1.0 AU for STEREO-A, Wind and SOHO . This results in asymmetric uncertainties in the path length acquired by VDA. The errors in the TSA injection times are estimated based on the uncertainties obtained due to the asymmetric error bars in the path length obtained from VDA, where the earliest possible emission time at the Sun corresponds to the longest possible path length that VDA yields, and the latest possible emission time corresponds to the particles traveling a completely straight path.The resulting parameters are provided in Table~\ref{tab:table1}, in which the path length $L$ is provided in the last column.}

{We find that the first electrons to reach STEREO-A at 1~AU are intermediate (225--255~keV) and high energy electrons (0.7--1.4~MeV) at 11:34 and 11:35~UT, respectively. These are shortly followed by near-relativistic electrons (55--65~keV) that reach the spacecraft at 11:38~UT. The Wind spacecraft observes energetic electrons at a later time, indicating that STEREO-A established an earlier connection to the injection region. The inferred injections times (Table~\ref{tab:table1}) suggest that the lower energy electrons (55--65~keV) are the first to be injected along open field lines, followed by the intermediate energy electrons that were injected $\sim$7~minutes later and high energy electrons $\sim$10~minutes later. The low-energy electrons at STEREO-A have an injection time of 11:15$^{+4}_{-5}$~minutes, which is close to the onset time of all radio bursts observed, while the high-energy electrons are closer in time to the herringbone sources that start to travel across the central meridian ($\sim$11:20~UT). The type II and type III radio bursts occur at a similar time. However, based on the magnetic field extrapolations, the type II propagates in a region of exclusively closed field lines forming a coronal streamer and thus the type II electrons are unlikely to escape the low corona. The type II also propagates in a north-westerly direction where there is no magnetic connectivity to STEREO-A (see Fig.~\ref{fig:fig8}). Regarding the type III bursts, there are no imaging observations available at low frequencies and it is not possible to determine their propagation directions. The electron beams generating herringbone bursts, however, propagate in a direction that meets open field lines which are likely to connect to STEREO-A (see Fig.~\ref{fig:fig8}).}

\begin{figure}[ht]
\centering
    \includegraphics[width=0.9\linewidth, trim = 0cm 0cm 0cm 0cm]{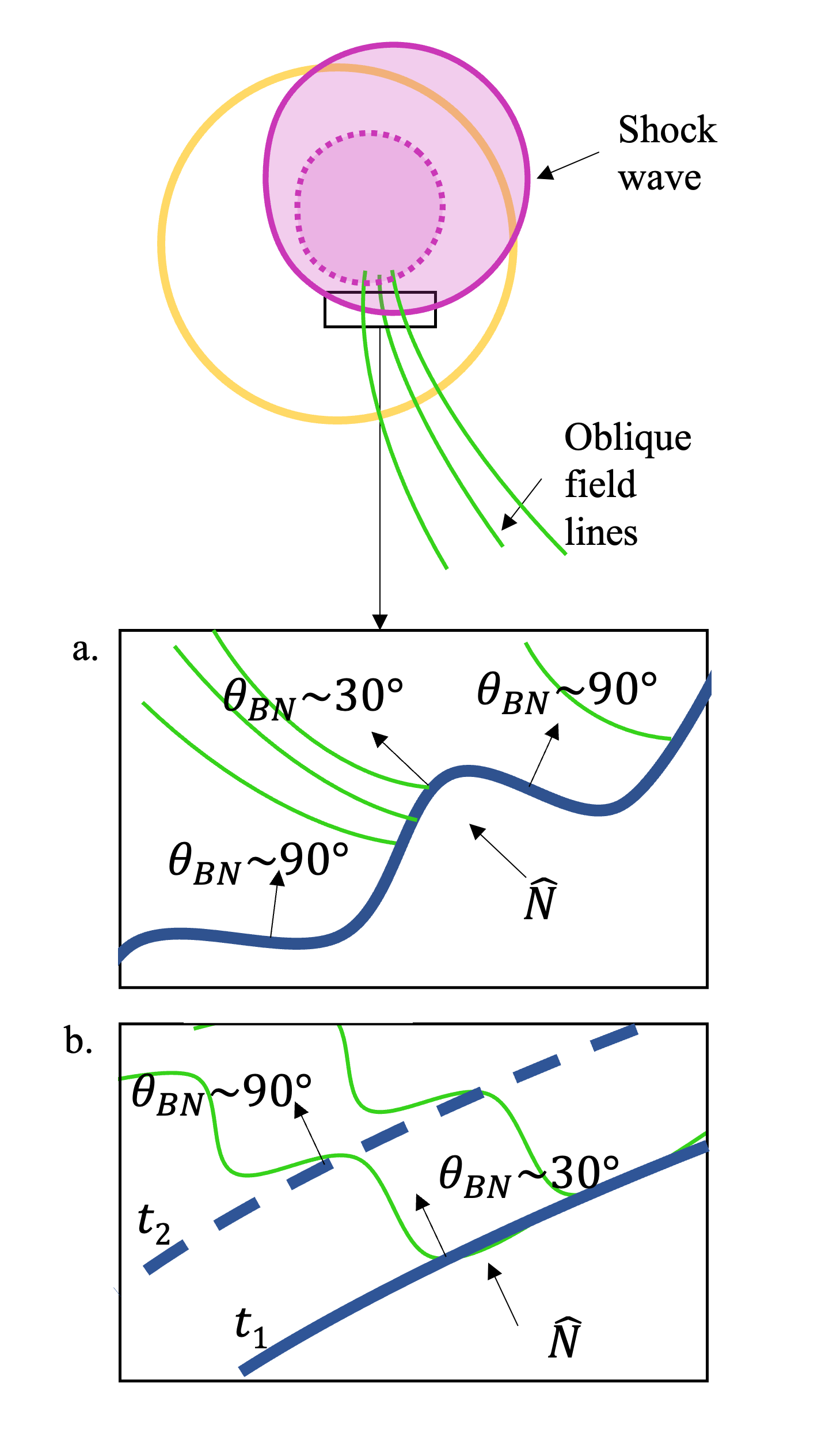}
    \caption{Cartoon showing the two possible scenarios where a globally oblique or quasi-parallel shock can behave as a quasi-perpendicular shock locally. a. A rippled shock front encountering oblique field lines. b. A planar shock front encountering non-ideal magnetic field lines at two consecutive times: $t_1$ and $t_2$.}
    \label{fig:fig11}
\end{figure}


\section{Discussion} \label{sec:discussion}

{We presented the first observational study where imaging of radio shock signatures in the low corona are compared to the occurrence of energetic electrons observed at spacecraft. We also presented rare radio imaging observations of herringbone bursts and show that they follow the expansion of a shock wave and occur at a different location to typical type II emission lanes. Our results show that the propagation directions of the type II burst and herringbone radio sources seem to follow the regions in the corona where certain conditions are met for the CME to drive a shock. For example, in these regions the CME drives a wave that is faster than the local Alfv\'en and magnetosonic speeds and forms a shock. The shock wave is also quasi-perpendicular in the case of the type II burst.} 

{We found two distinct regions where electrons are accelerated in the low corona, and we also found differences between the radio emission generated by these electrons. The first region is in a north-westerly direction where the shock is accompanied by a typical type II burst drifting to lower frequencies with fundamental and harmonic lanes and band-splitting. The type II burst has a curved trajectory that is cospatial with the path outlined by a high density ridge with low Alfv\'en and magnetosonic speeds provided by the MHD model. Here, the type II is also cospatial with a narrow region where the shock is close to $\theta_{BN}\sim90^{\circ}$, as expected from previous radio observations combined with MHD simulations \citep[e.g.,][]{jebaraj2021, kouloumvakos2021}.}

{The second region of electron acceleration is in a southwards direction where the radio emission shows emission lanes composed of herringbones with a slower drift rate than the preceding type II burst. The herringbone propagation direction also appears to be influenced by a narrow elevated density ridge in the corona characterised by low Alfv\'en and magnetosonic speeds. Unlike the type II burst, the herringbones are emitted near a region where the shock is instead oblique $\theta_{BN}\sim30^{\circ}$. This is likely explained by the fact that global parallel shocks can behave as perpendicular shocks locally for one of the following reasons: the existence of non-planar shock fronts or the upstream magnetic field has large-amplitude fluctuations. A schematic of these two possible scenarios is presented in Fig.~\ref{fig:fig11}. A non-planar shock front, such as a wavy or rippled shock \citep[e.g.,][]{zl93}, can switch between quasi-parallel and quasi-perpendicular as the shock ripples across magnetic field lines (see Fig.~\ref{fig:fig11}a). At small scales different portions of the shock ripples can be quasi-perpendicular to the magnetic field. Such a shock front also has the potential of creating magnetic traps that can aid in re-energisation of electrons due to multiple shock crossings \citep[][]{magdalenic02, morosan2020c} that can then escape in opposite directions from the shock and generate opposite drifting herringbone bursts. One possible cause for a rippled shock front is that, in the case of a strong quasi-parallel shock, the ramp is formed as a result of interactions between various energy dissipation processes which drive waves upstream leading to local non-linear steepening. These processes are quasi-periodic and can lead to ripples where locally the shock front is quasi-perpendicular \citep[see for example][]{Krasnoselskikh2002, lembege2004}. If we consider the shock lateral expansion speed of 830~km/s at 11:30~UT and the low Alfv\'en speed in the region the shock propagates into ($\sim$100~km/s), we obtain a high Alfv\'enic Mach number of $\sim$8 which indicates the presence of a strong shock during the time of herringbone emission. The second option is that the upstream magnetic field has large amplitude fluctuations. The shock geometry then changes continuously as the shock expands (see Fig.~\ref{fig:fig11}b). Large amplitude magnetic field structures in parallel shocks have been suggested before to explain herringbone emission \citep[][]{mann1995}. Shock wave interaction with non-ideal field lines have also been suggested in a similar study by \citet{morosan2020c}, where the existence of herringbone-like emissions with opposite drift rates were explained by a shock wave encountering helical quasi-open field lines forming the legs of an earlier CME. Thus, a global parallel or oblique shock may not necessarily indicate that the generation of type II and herringbone bursts is absent \citep[e.g.,][]{mann1995}. The small scale properties of the shock front are likely important in the generation of herringbone bursts. We note that the shock rippling in combination with high density and low Alfv\'en speed regions in the corona are the most likely factors that facilitate the presence herringbone emission. Only two such regions are immediately obvious in the MHD model in the vicinity of the eruption and these already coincide to the locations of type II and herringbone centroids. }

{In this event, the inferred injection time (11:15$\pm$5~UT) of near-relativistic electrons (55--65~keV) observed by the well-connected STEREO-A spacecraft is close to the onset time of the radio emissions: type II (11:15), type III (11:16) and herringbones (11:17). The herringbone electrons, however, are the most likely to encounter open magnetic field lines as they propagate southwards with the shock expansion as indicated by the MAST model. The type II and type III bursts occur 1 to 2~minutes earlier than the herringbones, however, the type II propagates inside a region of exclusively closed field lines forming a coronal streamer. The presence of type III bursts with a similar onset time cannot leave out the possibility that some of the in situ electrons are flare-accelerated. However, the type III bursts are emitted over a shorter period of time compared to the herringbone event. There is also a possibility that some of the in situ electrons are related to the second episode of HXRs since they also have a common emission time at the Sun (11:15~UT). In-depth HXR observations of the eruption with STIX \citep{purkhart2023} show the presence of a HXR source at the edge of a coronal dimming that is believed to mark one of the footpoints of the eruption. However, the footpoint near the coronal dimming only appears at a later time during the last identified HXR peak of the event (after 11:37~UT; time at the Sun 11:34~UT). This is later than the inferred injection time of STEREO-A electrons (11:15~UT). In addition, the derived electron flux from the STIX spectrum is at its highest during the first HXR peak after which it gradually drops \citep{purkhart2023}. The first HXR peak occurs a few minutes before the inferred injection time of electrons at STEREO-A (see Table~\ref{tab:table1}). The acceleration regions of HXR electrons may still have a connection to the in situ electrons due to the good magnetic connectivity of the eruption to STEREO-A but possibly not so well connected to the first arriving electrons. We also note that no clear indication of Langmuir waves is observed at STEREO-A and Wind. The presence of Langmuir waves has been attributed in past studies to a good magnetic connectivity to the acceleration site \citep[e.g.,][]{gomez2021}. However, this may not always be the case since the location of radio sources and CME propagation direction already show a good connectivity to STEREO-A.}

{The herringbone electrons have an energy with a lower limit of 8.4~keV, which is close to the energy range of the near-relativistic electrons that have a similar injection time. In addition, the electron beams which generate herringbone bursts propagate in a direction that intersects open field lines that are most likely magnetically connected to STEREO-A. Thus, an escape path of near-relativistic electrons is possible if they are emitted in the same location as the herringbone radio sources. The coincidental injection times and the spatial locations of herringbones and open field lines that can connect to STEREO-A, indicate that if the near-relativistic electrons observed at this spacecraft are shock-accelerated, then the region where the shock accelerates the herringbone producing electrons is also the only region where the in situ low energy electrons can be produced. The shock-acceleration scenario of the in situ electrons is also supported by Fig.~\ref{fig:fig10} where the high anisotropies seen later at Wind also suggest a direct connection to the source, in addition to STEREO-A. Compared to STEREO-A energetic particles arrive at near-Earth spacecraft with a significant delay of up to $\sim$9~minutes. This delay is also present when comparing the inferred injection times between STEREO-A and near-Earth spacecraft with an inferred electron injection delay of $\sim$7~minutes. The reason for this delay is likely due to the fact that STEREO-A is better magnetically connected to the injection site. However, the significant electron anisotropy observed at Wind (see Fig.~\ref{fig:fig10}) suggests that Wind is also well-connected to the injection region but the connection was established about 9~minutes later than that of STEREO-A. This could be due to the time it takes for the shock to expand in the low corona leading to a later shock connection with near-Earth spacecraft. The open field lines connecting to Earth are located close to the western limb and also at low latitudes close to the solar equator based on Fig.~\ref{fig:fig8}. In our reconstructions at 11:30~UT, when the herringbone and STEREO-A electrons occur close in time, the CME and accompanying shock are small and a significant distance away from low latitudes near the western limb as viewed from the Earth (see Fig.~\ref{fig:fig7}). The shock only reaches this location in the reconstructions at 11:35~UT (time at the Sun $\sim$11:27~UT), when it has the biggest chance of intersecting Earth-connecting open field lines. This expansion can explain at least partly the $\sim$9~minute delay between Earth and STEREO-A and it's also indicative of a large-scale shock wave as the source of the in situ electrons  observed at Earth. }


\section{Conclusion} \label{sec:conclusion}

{With the newly launched fleet of spacecraft, in particular PSP, SolO and BepiColombo that can reach closer distances to the Sun, combined with new ground-based imaging observations, similar studies can make better connections between the low and high coronal energetic particles during the onset of fast CMEs. However, such events are likely very rare as they require good magnetic connectivity to spacecraft at closer distances to the Sun and simultaneous availability of radio imaging during large eruptive events. Radio imaging with other observatories such as the Low Frequency Array (LOFAR) will also allow to track the herringbone electrons to larger coronal heights which will aid in making better connections to the open coronal field. }


\begin{acknowledgements}{D.E.M acknowledges the Academy of Finland project `RadioCME' (grant number 333859) and Academy of Finland project 'SolShocks' (grant number 354409). J.P. acknowledges the Academy of Finland Project 343581. N.D.\ and I.C.J are grateful for support from the Academy of Finland (SHOCKSEE, grant No.\ 346902). E.K.J.K. and E.A. acknowledge the European Research Council (ERC) under the European Union's Horizon 2020 Research and Innovation Programme Project SolMAG 724391 and E.K.J.K. also acknowledges the Academy of Finland Project 310445. EA also acknowledges support from the Academy of Finland/Research Council of Finland (Academy Research Fellow grant number 355659). A.K. and D.E.M. acknowledge the University of Helsinki Three-Year Grant. All authors acknowledge the Finnish Centre of Excellence in Research of Sustainable Space (Academy of Finland grant numbers 312390, 312357, 312351  and 336809). This study has received funding from the European Union’s Horizon 2020 research and innovation programme under grant agreement No.\ 101004159 (SERPENTINE). We thank the radio monitoring service at LESIA (Observatoire de Paris) to provide value-added data that have been used for this study. We also thank the Radio Solar Database service at LESIA / USN (Observatoire de Paris) for making the NRH and ORFEES data available. We thank the eCALLISTO network for the continuous availability of radio spectra.}\end{acknowledgements}

\bibliographystyle{aa} 
\bibliography{Bib_AA.bib} 

\newpage

\begin{appendix} 

\section{Radio burst fluxes} \label{app:a}

   \begin{figure}[ht]
   \centering
          \includegraphics[width=\linewidth, trim = 0cm 0cm 0cm 0cm]{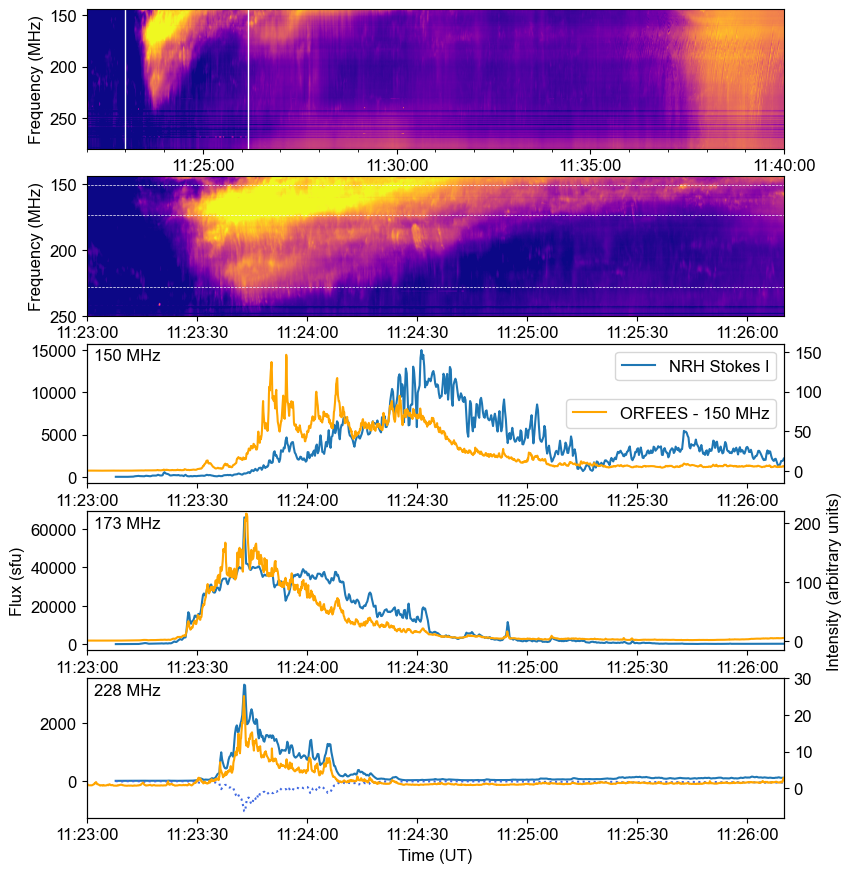}
      \caption{Zoomed-in dynamic spectrum of the type II burst and its flux density over time obtained from NRH images. The top panel shows the ORFEES dynamic spectrum during the type II harmonic lane. The last three panels show the flux of the type II bands together with the relative intensity extracted from the ORFEES dynamic spectrum at two frequencies: 228, 173 and 150~MHz. }
         \label{figA1}
   \end{figure}

   \begin{figure}[ht]
   \centering
          \includegraphics[width=\linewidth, trim = 0cm 0cm 0cm 0cm]{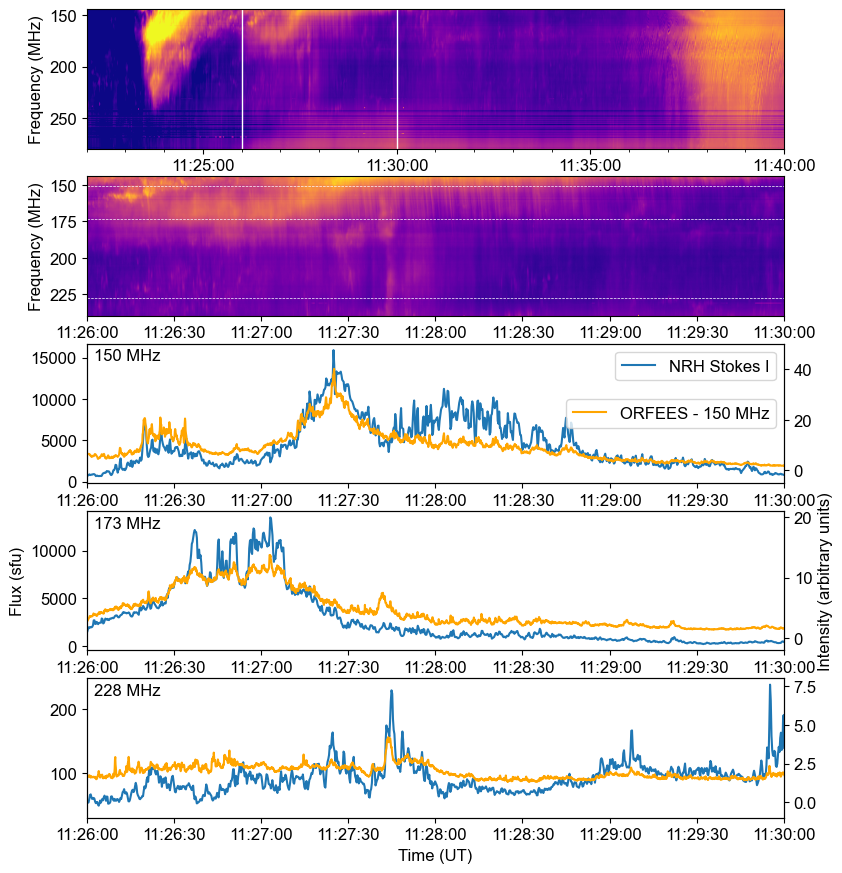}
      \caption{Zoomed-in dynamic spectrum of the herringbones bursts and their flux density over time obtained from NRH images. The top panel shows the ORFEES dynamic spectrum during the herringbone emission. The last three panels show the flux of herringbones together with the relative intensity extracted from the ORFEES dynamic spectrum at three frequencies: 228, 173 and 150~MHz. }
         \label{figA2}
   \end{figure}

{The flux of the radio sources imaged by the NRH at 150, 173 and 228~MHz is compared to the relative intensity of the type II lanes and herrignbones in the ORFEES dynamic spectrum as shown in Figs.~\ref{figA1} and \ref{figA2}. In these Fig.s, the flux in the NRH images is estimated by adding the pixels over the extent of the radio source which is defined as the pixels with values above 20\% of the maximum intensity level in each image. This flux is then compared with the relative intensity of the Type II and herringbone bursts obtained from the ORFEES dynamic spectra. The ORFEES spectra have a lower temporal resolution than the NRH images, 100~ms compared to 250~ms, respectively. The bottom three panels of Figs.~\ref{figA1} and \ref{figA2} show the NRH Stokes I (solid blue line) fluxes together with the ORFEES relative intensity of the type II and herringbone bursts at 228, 173 and 150~MHz. These Fig.s show an overall good association between the imaged radio source and the relative intensity peaks of the type II and herringbone bursts obtained from the dynamic spectra. This is a confirmation that the imaged bursts represent the type IIs and herringbones observed.  }

\section{Expansion speeds of the shock} \label{app:b}

\begin{figure}[ht]
   \centering
          \includegraphics[width=\linewidth, trim = 0cm 0cm 0cm 0cm]{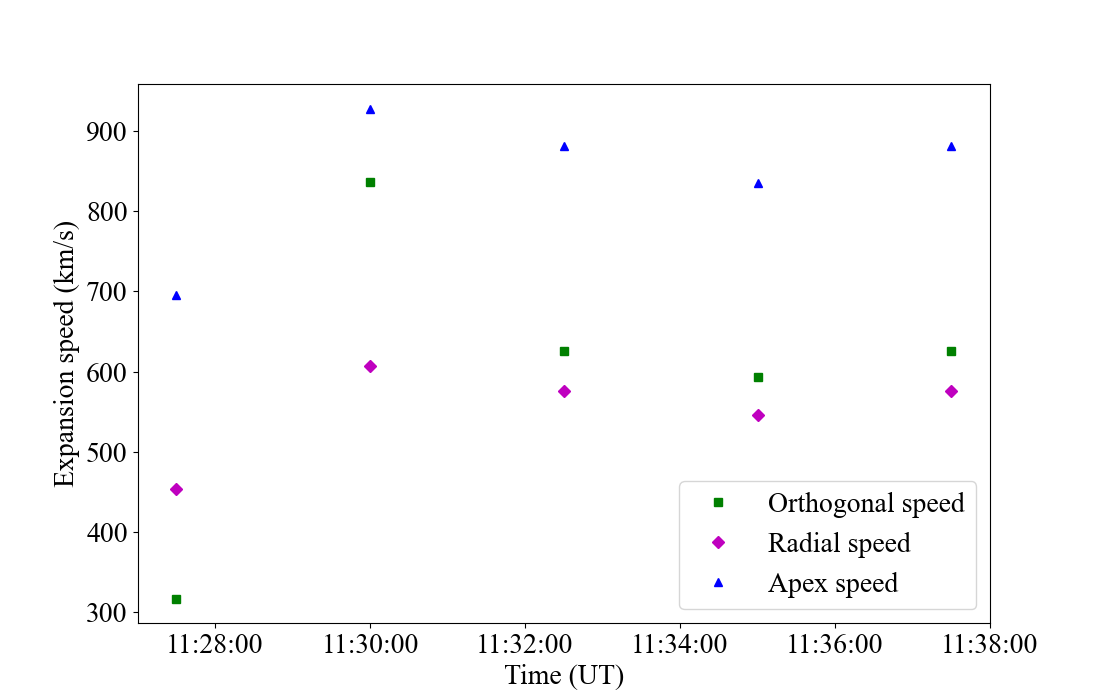}
      \caption{The EUV shock speeds along the apex, radial and orthogonal axes of the sphereoid fitting estimated over time. }
         \label{figB1}
   \end{figure}

{The shock reconstruction was applied to six time steps on 28 March 2022, while the shock is visible in EUV images. Based on the fitted values of the apex height as well as the radial and orthogonal axes, the EUV shock speeds in these directions can be estimated over time. The shock speeds during the six time steps of the reconstruction are shown in Fig.~\ref{figB1}.}

\section{Velocity dispersion analysis} \label{app:b}

\begin{figure*}[ht]
   \centering
          \includegraphics[width=0.7\linewidth, trim = 0cm 0cm 0cm 0cm]{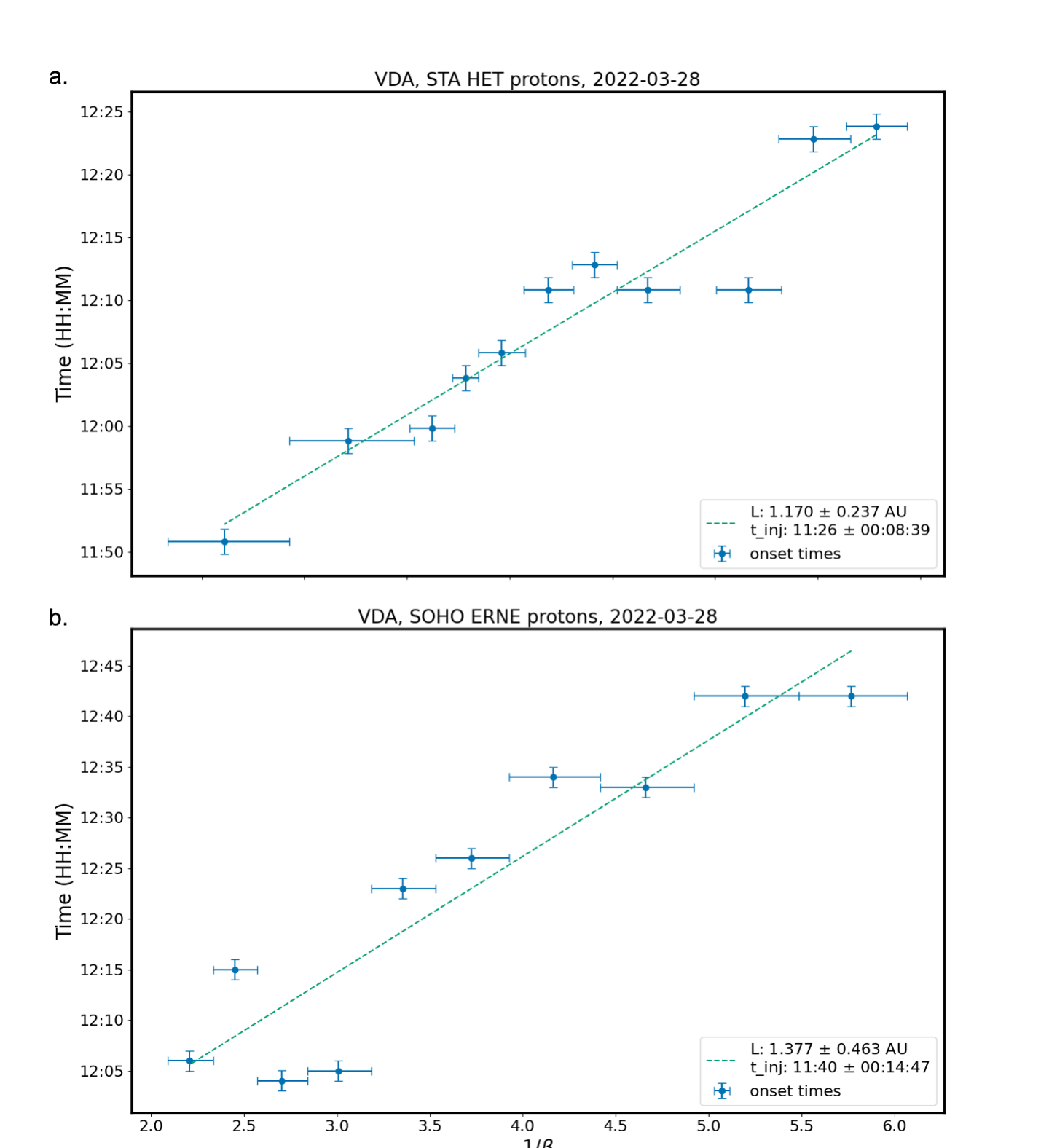}
      \caption{Results of the proton VDA analysis using STEREO-A/HET (a) and SOHO/ERNE (b). }
         \label{figC1}
   \end{figure*}

{To infer the solar injection times of SEPs, we employed the VDA model assuming a common injection time and propagation path length of all particles. We performed a VDA for protons using STEREO-A/HET and SOHO/ERNE observations to provide the path length and injection time of protons accompanied by error bars. The path length was then used in the TSA method to obtain the inferred injection time of electrons, since VDA for electrons yielded an unphysical path length. The results of the proton VDA analysis is shown in Fig.~\ref{figC1}.}

\end{appendix} 

\end{document}